\documentclass[twocolumn,aps,showpacs,preprintnumbers,amsmath,amssymb,10pt]{revtex4-1}

\usepackage{graphicx}
\usepackage{dcolumn}
\usepackage{bm}

\begin{document}

\title{Sub-barrier enhancement of fusion as compared to a microscopic method in $^{18}$O+$^{12}$C}

\author{T.~K. Steinbach}
\author{J. Vadas}
\author{J. Schmidt}
\author{C. Haycraft}
\author{S. Hudan}
\author{R.~T. deSouza}
\email{desouza@indiana.edu}
\affiliation{%
Department of Chemistry and Center for Exploration of Energy and Matter, Indiana University\\
2401 Milo B. Sampson Lane, Bloomington, Indiana 47408, USA}%

\author{L.~T. Baby}
\author{S.~A. Kuvin}
\author{I. Wiedenh\"{o}ver}
\affiliation{
Department of Physics, Florida State University, Tallahassee, Florida 32306, USA}%

\author{A.~S. Umar}
\author{V.~E. Oberacker}
\affiliation{Department of Physics and Astronomy, Vanderbilt University, Nashville, Tennessee 37235, USA}%

\date{\today}

\begin{abstract}
\begin{description}
\item[Background] Measurement of the energy dependence of the fusion cross-section at sub-barrier energies provides an important test for theoretical models of fusion.
\item[Purpose] To extend the measurement of fusion cross-sections in the sub-barrier domain for the $^{18}$O+$^{12}$C system. Use the new experimental data to confront 
microscopic calculations of fusion.
\item[Method] Evaporation residues produced in fusion of $^{18}$O ions with $^{12}$C target nuclei were detected with good geometric efficiency and identified by measuring their
energy and time-of-flight. Theoretical calculations with a density constrained time dependent Hartree-Fock (DC-TDHF) theory include for the first time the effect of pairing
on the fusion cross-section.
\item[Results] Comparison of the measured fusion excitation function with the predictions of the DC-TDHF calculations reveal that the
experimental data exhibits a smaller decrease in cross-section with decreasing energy than is theoretically predicted.
\item[Conclusion] The larger cross-sections observed at the lowest energies measured indicate a larger tunneling probability for the fusion process. This larger
probability can be associated with a smaller, narrower fusion barrier than presently included in the theoretical calculations.
\end{description}
\end{abstract}

\maketitle

Understanding the origin of the elements, namely where and how they are formed, is one of the grand challenges in science \cite{NSAC07}. 
Independent of where they are formed whether
in stellar interiors \cite{Penionzhkevich10, Carnelli14} or on earth \cite{Zagrebaev12, Back14, Yanez14}, nucleosynthesis involves nuclear fusion. 
An accurate description of fusion is thus key to 
understanding nucleosynthesis. In recent years, fusion reactions have also been proposed to be important in exotic astrophysical environments, for example 
triggering X-ray superbursts that originate in the crust of an accreting neutron star \cite{Strohmayer06}. The putative reactions involve fusion of neutron-rich 
light nuclei at sub-barrier energies \cite{Horowitz08, Horowitz09a, Horowitz09b}. Initial measurements of fusion induced with neutron-rich light nuclei suggest 
an enhancement of the fusion probability as compared to standard models of fusion-evaporation \cite{Rudolph12}. At sub-barrier energies one is particularly 
sensitive to the microscopic degrees-of-freedom as the two nuclei collide, hence a microscopic treatment is the most relevant. For reactions at sub-barrier 
energies, the initial interpenetration of the matter distributions of the two nuclei is small but at the inner turning-point
the nuclei have significant overlap. 
Consequently, a theoretical approach that can accurately describe the matter distributions at the outer and inner turning-points is desirable.
The time-dependent Hartree-Fock (TDHF) theory provides a practical foundation for a fully 
microscopic theory of large amplitude collective motion and is thus well suited to study low-energy fusion reactions~\cite{Negele82, Simenel12}. 
Recent advances 
provide a more realistic treatment of tunneling through the use of microscopically obtained heavy-ion potentials~\cite{Umar06d, Washiyama08}.
To test the accuracy of these microscopic 
calculations, high quality experimental data is needed. In the present work, we present a measurement of the total fusion cross-section for $^{18}$O + $^{12}$C 
that extends one order of magnitude lower in cross-section than previous measurements. We utilize this data to compare to
the microscopic calculations of fusion.

\begin{figure}[!htb]
\includegraphics[width=8.6cm]{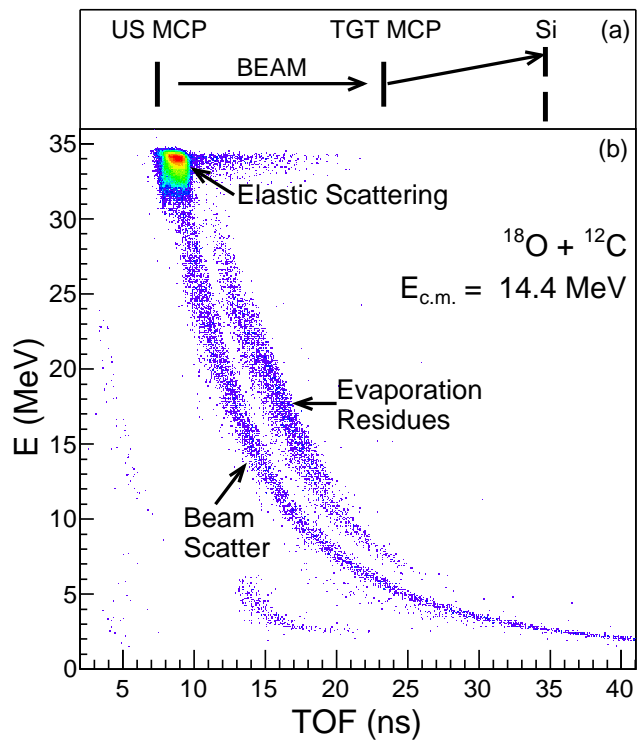}
\caption{\label{fig:etof} (Color online) Top panel: Schematic illustration of the experimental setup. 
Bottom panel: Energy versus time-of-flight spectrum for $^{18}$O ions incident on $^{12}$C target nuclei at E$_{\mathrm{c.m.}}$=14.4~MeV.
Color is used to represent yield in the two dimensional spectrum on a logarithmic scale.}
\end{figure}

To conduct the experiment, a beam of $^{18}$O ions with an intensity of 1.5-4$\times$10$^5$~p/s was accelerated to energies between 
E$_{\mathrm{lab}}$ = 16.25~MeV  and E$_{\mathrm{lab}}$ = 36~MeV using the FN tandem at Florida State University's John D. Fox accelerator center. 
The beam was pulsed at a frequency of 12.125~MHz. As the beam energy is critical to the accurate measurement of the fusion excitation function, 
the accuracy of the accelerator energy calibration was checked using known proton resonance
energies and determined to be within 7~keV. After passing through 
a microchannel plate (MCP) detector approximately 1 m upstream of the target (US MCP), the beam was incident on a 93 $\mu$g/cm$^2$ thick $^{12}$C foil which 
served as the target (TGT MCP) as depicted in Fig.~\ref{fig:etof}a. 
The $^{12}$C target foil also served as a secondary emission foil for a microchannel plate detector \cite{Steinbach14} thus providing a 
timing signal for a time-of-flight (TOF) measurement. Measurement of the TOF between the two MCPs allows one to reject beam particles scattered or degraded 
prior to the target. 

Earlier measurements of sub-barrier fusion for similar systems identified fusion events by measuring the $\gamma$-rays emitted by the fusion residues as they 
de-excited \cite{Cujec76, Christensen77}. While this approach, of tagging fusion through $\gamma$-rays, allows the use of thick targets 
and high intensity beams, it suffers from the low efficiency inherent 
in $\gamma$-ray detection and any uncertainties in the knowledge of the relevant decay channels. We have therefore elected to directly measure the fusion 
residues in order to determine the fusion cross-section. 

Reaction products were detected in the angular range 4.3$^\circ$ $\le$ $\theta_{\mathrm{lab}}$ $\le$ 11.0$^\circ$ using a segmented, annular silicon 
detector which provided both an energy and fast timing signal \cite{deSouza11}. Due to the kinematics of the reaction, the angular range subtended by 
this detector resulted in a high geometric efficiency for detection of fusion residues. Reaction products were distinguished on the basis of their energy and 
time-of-flight (ETOF) \cite{Steinbach14}. A typical ETOF spectrum measured is depicted in Fig.~\ref{fig:etof}b where the energy corresponds to the energy 
deposited in the silicon detector while the time-of-flight is the time difference between the target MCP and the silicon detector.

The most prominent feature in Fig.~\ref{fig:etof}b is the peak at E $\approx$ 34~MeV that corresponds to elastically scattered particles. 
Originating from this peak is a locus of points with lower energies and longer TOF values. Points in this locus are scattered beam particles. 
Visible at larger TOF and clearly separated from the beam scatter line is an island of reaction products. This island is populated by evaporation 
residues that result from fusion of the projectile and target nuclei to form a compound nucleus which subsequently de-excites. Protons and alpha 
particles which are emitted during this de-excitation cascade of the compound nucleus manifest themselves in the spectrum with a characteristic energy 
time-of-flight relationship. Alpha particles are observed with energies between 10~MeV and 25~MeV and TOF values of approximately 5~ns. Protons are observed 
at deposited energies of E $<$ 6~MeV, consistent with the Si detector thickness, and TOF values of approximately 15~ns. The larger TOF values
observed for protons as compared to the alpha particles is due to the slower risetime exhibited by protons and the leading edge discrimination employed.
Also visible in the spectrum is a tail on the elastic peak which is constant in energy and extends to larger TOF values. This tail occurs with low probability 
(0.4$\%$) as compared to the elastic peak.

The measured evaporation residue cross-section was ascertained by using the measured number of beam particles incident on the target, the measured 
number of evaporation residues, and the known target thickness. The total number of beam particles incident on the target was determined by 
counting the coincidences between the MCP at the target position and the upstream MCP. The number of residues 
detected was established by selecting the appropriate region of the ETOF spectrum and summing the number of evaporation residues contained within 
it. The limits of the region of integration were established by calculating the TOF for different mass residues and using the beam scatter line 
as a reference. After accounting for the finite time resolution, an interval in mass number, 22 $\le$ A $\le$ 30 was used for measurements at E$_{\mathrm{c.m.}}$ $>$ 7.5~MeV 
and 24 $\le$ A $\le$ 30 for E$_{\mathrm{c.m.}}$ $\le$ 7.5~MeV. 

In order to determine the total fusion cross-section it is necessary to know the geometric efficiency of the experimental setup. The efficiency was 
determined by using a statistical model, evapOR \cite{evapOR}, which simulates the decay of a compound nucleus using a Hauser-Feshbach approach. 
By calculating the fraction of the evaporation residues that lie within the detector acceptance, the geometric efficiency of the experimental setup is obtained. 
The bombarding energy dependent  efficiency lies between 50$\%$ and 59$\%$. Using the efficiency together with the measured evaporation residue 
cross-section, the total fusion cross-section is extracted. Since the MCP efficiency affects both the counting of the total number of beam particles and the 
number of evaporation residues, it does not impact the measured total fusion cross-section.

\begin{figure}[!htb]
\includegraphics[width=8.6cm]{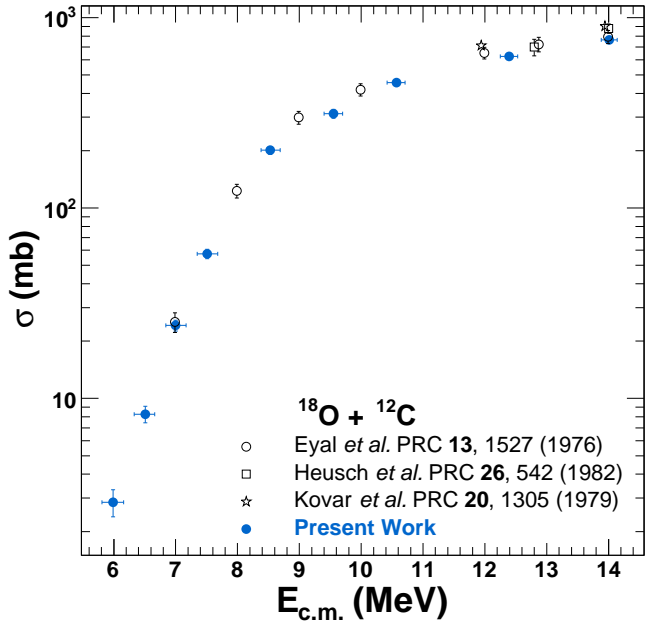}
\caption{\label{fig:xsect_exp} (Color online) The fusion excitation function for $^{18}$O + $^{12}$C is shown at energies near and below the Coulomb barrier. 
Literature values are shown as open circles \cite{Eyal76}, open squares \cite{Heusch82}, and open stars \cite{Kovar79} with the present data represented by 
solid circles.}
\end{figure}

The measured excitation function is displayed in Fig.~\ref{fig:xsect_exp} along with previously published results \cite{Eyal76, Kovar79, Heusch82}. 
As expected, the fusion cross-sections decrease with decreasing E$_{\mathrm{c.m.}}$ indicative of a barrier controlled phenomenon. 
It is noteworthy that even for the lowest energies measured an exponential decrease of the cross-section with decreasing energy is observed.
Vertical error bars on 
the present data include both the statistical uncertainties as well as a 2$\%$ systematic error. 
This systematic error is associated with the 
analysis. Horizontal error bars represent the uncertainty in whether the fusion occurs at the front or back of the target foil. Using the direct 
measurement of evaporation residues as done in the present experiment, previous measurements only measured the fusion cross-section down to the 
25~mb level \cite{Eyal76}. In contrast, in the present work we measure the fusion cross-section down to the 2.8~mb level, close to a full order of magnitude 
lower in cross-section. At energies where the present dataset overlaps with existing data, overall agreement of the cross-sections is good, 
close to the statistical uncertainties of the prior measurements. This overall agreement not only indicates that our approach in extracting the fusion 
cross-section is sound but that there are no significant uncertainties in the values of the target thickness or detector efficiency. Closer comparison of the 
present dataset with the data of Ref.~\cite{Eyal76} indicates that the presently measured cross-sections are approximately 7$\%$ lower for 
E$_{\mathrm{c.m.}}$ $\ge$ 10~MeV. While the prior measurements required integration of an angular distribution measured with a low geometric efficiency 
detector, the present measurement directly measures a large fraction of the evaporation residue yield. We therefore believe that the present cross-sections 
are more accurate. It should be noted that the statistical quality of the present dataset is substantially better than that of the earlier measurements.

In recent years it has become possible to perform TDHF calculations on a 3D Cartesian grid thus not 
requiring any artificial symmetry restrictions and with 
much more accurate numerical methods~\cite{umar1991a,Umar06a,maruhn2014}. In addition, the quality of the effective interactions has been substantially 
improved~\cite{reinhard1988, umar1989, Chabanat98a, Klupfel09a, Kortelainen10}.
Over the past several years, the density constrained TDHF (DC-TDHF) method for calculating heavy-ion potentials~\cite{Umar06d} has been 
employed to calculate heavy-ion fusion cross-sections with remarkable success~\cite{Keser12,Back14}.
While most applications have been for systems involving heavy nuclei, recently the theory was used to study above and below barrier
fusion cross-sections for lighter systems, specifically for reactions involving various isotopes of O+O and O+C~\cite{Umar12,deSouza13} relevant
for the reactions that occur in the neutron star crust.
One general characteristic of TDHF and DC-TDHF calculations for light systems is that the fusion cross-section at energies 
well above the barrier are
usually overestimated~\cite{simenel2013a,umar2014a}, whereas an excellent agreement is found for sub-barrier cross-sections~\cite{Umar12}.
This is believed to be due to various breakup channels in higher energy reactions of these lighter systems that are not properly
accounted for in TDHF dynamics and contribute to fusion instead.
Nevertheless, the agreement is remarkable given the fact that the only input in DC-TDHF is the Skyrme effective N-N interaction, 
and there are no adjustable parameters.

An unfortunate consequence of the TDHF approach, however, is the inability to treat pairing during the collision process. This shortcoming in the 
inclusion of pairing has led to the prediction of deformation of the ground state for some even-even nuclei such as $^{18,20}$O in disagreement with 
self-consistent mean field calculations that include pairing. To overcome this shortcoming, in prior work an average of all 
orientations of the deformed nucleus 
with respect to the target nucleus has been performed~\cite{deSouza13}. 
It can be qualitatively argued that this averaging nonetheless results in a 
larger fusion cross-section as compared to the spherical nucleus.

\begin{figure}[!htb]
\includegraphics[width=8.6cm]{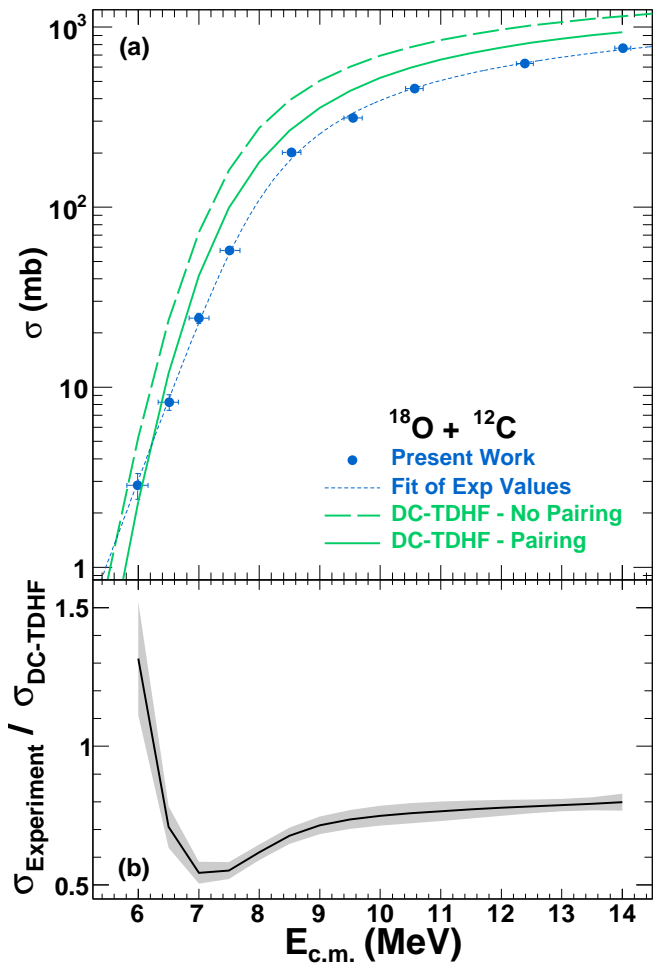}
\caption{\label{fig:xsect_ratio} (Color online) Top panel: Comparison of the experimentally measured fusion cross-sections (closed symbols) with the results of
the DC-TDHF calculations with (solid line) and without (dashed line) pairing. Also shown, as a dotted line, is the result of a fit to the experimental data (see text for details).
Bottom panel: Energy dependence of the ratio of the experimentally measured cross-sections to the DC-TDHF predictions which include pairing. The shaded band depicts the
uncertainty in the ratio due to the uncertainty in the experimental cross-sections.}
\end{figure}

In this work we report for the first time on the inclusion of pairing in the DC-TDHF calculations.
This was achieved by performing a BCS pairing calculation for the static solution of $^{18}$O resulting in a spherical
nucleus with a subsequent density constraint calculation to produce this density as a solution of the ordinary
Hartree-Fock equations in the spirit of the density-functional theory. This nucleus with frozen occupations was then
used in the TDHF time evolution. Subsequent density-constraint calculations in DC-TDHF method preserves this
spherical shape during the entrance channel dynamics.
As can be seen in Fig.~\ref{fig:xsect_ratio}a, inclusion of pairing in the DC-TDHF calculation for $^{18}$O + $^{12}$C results in a significant 
reduction of the fusion cross-section. The standard DC-TDHF calculations are presented as the dashed curve while the calculations that include 
pairing are depicted as the solid curve. At all energies, pairing acts to reduce the fusion cross-section. 
At the highest energies shown pairing reduces the cross-section to $\approx$~80$\%$ of the value calculated without pairing. This difference between the 
calculations with and without pairing increases dramatically as the incident energy decreases. At the lowest energies shown the introduction of pairing in 
the calculation reduces the cross-section to $\approx$~36$\%$ of the cross-section calculated without pairing.

Comparison of the experimental fusion excitation function with the DC-TDHF microscopic calculations is presented in Fig.~\ref{fig:xsect_ratio}a. 
The presently measured fusion cross-sections, previously shown in Fig. \ref{fig:xsect_exp}, are
indicated as solid symbols. Overall comparison of the experimental cross-sections with the DC-TDHF calculations indicate that the experimental 
cross-sections are lower than the theoretical predictions. In order to facilitate a quantitative comparison of the experimental excitation function with 
the theoretical predictions, we have fit the experimental cross-sections with a functional form \cite{Wong73} that describes the penetration 
of an inverted parabolic barrier:
\begin{equation}
\sigma = \frac{R_c^2}{2E}\hbar\omega \cdot ln \left \{ 1+exp\left [\frac{2\pi}{\hbar\omega}(E-V)\right] \right \}
\end{equation} 
where $R_c$ is the radius of the fusion barrier, $V$ is the height of the interaction barrier, $\omega$ is the frequency, and $E$ is the incident energy.
The best fit achieved using this functional form is shown as the dotted line in Fig.~\ref{fig:xsect_ratio}a and has values of $R_c$ = 7.24 $\pm$ 0.16 fm,
$V$ = 7.62 $\pm$ 0.14~MeV, and $\hbar\omega$ = 2.78 $\pm$ 0.29~MeV, remarkably the calculated DC-TDHF barrier also has a barrier height
of 7.60~MeV. Shown in Fig.~\ref{fig:xsect_ratio}b is 
the ratio of the fit of the experimentally measured cross-sections to the DC-TDHF calculations with pairing. For energies E$_{\mathrm{c.m.}}$ $>$ 9~MeV, 
the ratio $\sigma_{\mathrm{Experiment}}$/$\sigma_{\mathrm{DC-TDHF}}$ is $\approx$ 0.75 and decreases weakly with decreasing energy. 
A stronger decrease in the ratio is observed as the energy decreases from E$_{\mathrm{c.m.}}$ = 9~MeV to E$_{\mathrm{c.m.}}$ = 7~MeV. At this energy, 
the ratio is minimum with a value of 0.54. As the incident energy decreases further, the ratio increases reaching a value of 1.32 at the lowest energy measured,
E$_{\mathrm{c.m.}}$ $=$ 6~MeV. 
The presence of breakup reactions at energies above the barrier could explain 
the fact that the ratio is less than unity in this energy range. 
With decreasing incident energy, the role of breakup reactions diminishes hence the ability of 
the DC-TDHF method to describe fusion is expected to improve. We therefore focus our attention on the comparison of the model and experiment in the sub-barrier region.
The key feature in the ratio is therefore its change with decreasing incident energy in the sub-barrier domain, specifically its increase from a value smaller than unity to 
a value larger than unity. This trend emphasizes that the experimental and theoretical excitation functions have different shapes with the experimental cross-section
falling more slowly with decreasing incident energy than is theoretically predicted by the DC-TDHF method. This enhancement of the experimental fusion cross-sections relative to 
the DC-TDHF predictions is a factor of $\approx$~2.4 as the incident energy decreases from E$_{\mathrm{c.m.}}$ = 7~MeV to E$_{\mathrm{c.m.}}$ = 6~MeV.
We have assessed the impact of the experimental uncertainties on the ratio presented and display 
the result as a shaded band in  Fig.~\ref{fig:xsect_ratio}b.
It is clearly evident that the trends exhibited by the ratio are significantly 
larger than the magnitude of the uncertainties.

The fact that the sub-barrier experimental fusion cross-sections decrease more slowly with decreasing energy than the calculated cross-sections 
can be interpreted as a larger tunneling probability for the experimental data as compared to the theoretical calculations. This enhanced tunneling probability can 
be associated with a narrower barrier, which deviates from an inverted parabolic shape. 
The fundamental reason that the barrier determined from the experimental data is weaker than in the theory is presently
unclear. 
It should also be recalled that within the DC-TDHF calculations, inclusion of pairing decreased the predicted cross-sections. It was assumed that the initial 
occupation numbers calculated 
with pairing were frozen as the reaction dynamics proceeded. It can be argued that relaxing this stringent condition would result in larger cross-sections. 
Unfortunately, microscopic calculations which allow the pairing to evolve in response to changes in the shape of the nuclear system 
as the fusion proceeds are beyond the scope of the present work. Such calculations would provide a more realistic treatment of the impact of pairing on fusion.
It is noteworthy that the previous experimental data \cite{Eyal76} only extended down to E$_{\mathrm{c.m.}}$ = 7~MeV. The dramatic increase in cross-section relative to the
DC-TDHF method occurs at energies below E$_{\mathrm{c.m.}}$ = 7~MeV. 
This enhancement of the fusion cross-section in the sub-barrier domain demonstrates the importance of measuring the sub-barrier fusion cross-section 
for light, heavy-ion reactions. This sub-barrier cross-section enhancement could yield new insight into the fusion dynamics of neutron-rich light nuclei.

We wish to acknowledge the support of the staff at Florida State University's John D. Fox accelerator in providing the high quality beam that 
made this experiment possible. This work was supported by the U.S. Department of Energy under Grant No. DE-FG02-88ER-40404 (Indiana University) 
and Grant No. DE-FG02-96ER40975 (Vanderbilt University) and the National Science Foundation
under Grant No. PHY-1064819 (Florida State University).

\bibliography{fusion_18O_v3}

\end{document}